\documentclass[preprintnumbers,superscriptaddress,showpacs,pra]{revtex4}
\usepackage{amsmath}
\usepackage{epsfig}
\usepackage{graphicx}
\usepackage{color}

\input{tcilatex}
\begin{document}

\title[curvature induced force]{Heisenberg equation for a nonrelativistic
particle on a hypersurface: from the centripetal force to a curvature
induced force}
\author{D. K. Lian, L. D. Hu, and Q. H. Liu}
\affiliation{School for Theoretical Physics, School of Physics and Electronics, Hunan
University, Changsha 410082, China}
\date{\today }

\begin{abstract}
In classical mechanics, a nonrelativistic particle constrained on an $N-1$
curved hypersurface embedded in $N$ flat space experiences the centripetal
force only. In quantum mechanics, the situation is totally different for the
presence of the geometric potential. We demonstrate that the motion of the
quantum particle is "driven" by not only the the centripetal force, but also
a curvature induced force proportional to the Laplacian of the mean
curvature, which is fundamental in the interface physics, causing curvature
driven interface evolution.
\end{abstract}

\maketitle

For an nonrelativistic particle constrained on ($N-1$)-dimensional smooth
curved surface $\Sigma ^{N-1}$ in flat space $R^{N}$ ($N\succeq 2$), there
is an apparent force, the centripetal force. It is a commonly hold belief
that in quantum mechanics, Heisenberg equation for the time derivative of
momentum takes the form identical to the that in classical mechanics \cite%
{weinberg}, c.f. Eq. (\ref{cff1}). In fact, the situation is far more
complicated than what is anticipated. This is because in quantum mechanics
for motion on the hypersurface, there is a curvature induced potential \cite%
{DeWitt,jk,dacosta,fc}\ that has no classical correspondence, and we can by
no mean assume that same form of the Ehrenfest theorem for the time
derivative of mean value of the momentum applies.

As to the curvature induced potential,\ it was first suggested by DeWitt in
1957 \cite{DeWitt}, but its identification was due to Jensen and Koppe \cite%
{jk} in 1971 and subsequently da Costa \cite{dacosta} during 1981-1982, who
developed a \textit{confining potential formalism} to deal with the free
motion on the curved surface. By the confining potential formalism we mean
to write the Schr\"{o}dinger equation within the uniform flat space $R^{N}$
within sufficiently high potential barriers on both sides of the surface,
and then squeeze the width of barriers. Since the energy difference between
the excited and the ground state of the particle along the direction normal
to the surface is very much larger than that of the particle along the
tangential direction so that the degree of freedom along the normal
direction is actually frozen to the ground state, an effective dynamics for
the constrained system on the surface is thus resulted in, which contains a
well-defined form of the curvature induced potential, \textit{the geometric
potential }as called later, c.f. Eq. (\ref{GP}). In 2013, Liu \cite{liu13-1}
applied the confining potential formalism to the momentum and demonstrated
that it became the geometric momentum \cite{liu11}, c.f. Eq. (\ref{GM}). The
confining potential formalism is now widely used to predict
curvature-induced consequences in two-dimensional curved surfaces or curved
wires \cite{packet1}. Experimental confirmations of the geometric potential
include an optical realization of the potential \cite{exp1} in 2010 and the
potential in a one-dimensional metallic $\mathit{C}_{60}$ polymer with an
uneven periodic peanut-shaped structure in 2012 \cite{exp2}. An interesting
application of geometric momentum is that the propagation of surface plasmon
polaritons on metallic wires is in 2015 found to be governed by two solely
curvature-induced geometric momenta, leading to a significant modification
of the waveguide dispersion, i.e. a change of their phase velocity \cite%
{waveguide}.

In the present study, we show that the time derivative of momentum in the
Heisenberg equation gives not only the centripetal force which has classical
correspondence, but also a curvature induced force proportional to the full
Laplacian of the mean curvature, of which the intrinsic Laplacian part in
interface physics is responsible for the curvature driven interface
evolution, and phase transition due to the diffusion, etc. \cite{packet2},
which has nevertheless no classical correspondence.

Let us consider the surface equation $f(x)=0$, where $f(x)$ is some smooth
function of position $\mathbf{x\equiv }(x_{1},x_{2},...x_{N})$ in $R^{N}$,
whose normal vector is $\mathbf{n}\equiv \nabla f(x)/|\nabla f(x)|$. We can
always choose the equation of the surface such that $|\nabla f(x)|=1$, so
that $\mathbf{n}\equiv \nabla f(x)$. No matter what form of the surface
equation we choose, only the unit normal vector and/or its derivatives enter
the physics equation. In classical mechanics, the equation of motion of the
particle on the surface is, \cite{weinberg,ikegami,liu16,liu17},%
\begin{equation}
\frac{d}{dt}\mathbf{p=}-\mathbf{n}\left( \frac{\mathbf{\mathbf{p}}\cdot 
\mathbf{\nabla \mathbf{n}}\cdot \mathbf{\mathbf{p}}}{\mu }\right) ,
\label{cff1}
\end{equation}%
where $\mu $ symbolizes the mass of the particle, and $d\mathbf{p}/dt$
denotes the derivative of the momentum $\mathbf{p}$ with respect to $t$. The
right hand side of this equation (\ref{cff1}) does not take the familiar
form of the centripetal force. This is because we do not consider the
geodesic the particle is bound to move along. With accounting for this fact,
Eq. (\ref{cff1}) becomes \cite{liu16},%
\begin{equation}
\frac{d}{dt}\mathbf{v=}-\mathbf{n}\frac{v^{2}}{R},  \label{cff2}
\end{equation}%
where $1/R$ stands for the first local curvature of the geodesic, and $%
\mathbf{v}$ represents the velocity of the particle. Our key finding is in
the following, in Heisenberg picture, the motion of equation can be given
by, 
\begin{equation}
\frac{d}{dt}\mathbf{p=}-\frac{1}{2}\left\{ \mathbf{n}\left( \frac{\mathbf{%
\mathbf{p}}\cdot \mathbf{\nabla \mathbf{n}}\cdot \mathbf{\mathbf{p}}}{\mu }%
\right) +\left( \frac{\mathbf{\mathbf{p}}\cdot \mathbf{\nabla \mathbf{n}}%
\cdot \mathbf{\mathbf{p}}}{\mu }\right) \mathbf{n}\right\} -\frac{\hbar ^{2}%
}{4\mu }\nabla ^{2}M\mathbf{n},  \label{gemforce}
\end{equation}%
where $M\equiv -n_{i,i}$ is the mean curvature of the surface and "$,i$" in
the subscript denotes the derivative with respect to the coordinate $x_{i}$,
and the $\nabla ^{2}\equiv \partial _{i}\partial _{i}$ in which hereafter
repeated indices are summed over. Operator $\mathbf{\mathbf{p}}\cdot \mathbf{%
\nabla \mathbf{n}}\cdot \mathbf{\mathbf{p}}%
=p_{i}n_{ij}p_{j}=p_{i}n_{ji}p_{j} $ is manifestly hermitian for $%
n_{ji}=n_{ii}$. Evidently, there is a curvature induced quantum force $f_{g}$%
,%
\begin{equation}
\mathbf{\chi }_{g}\equiv -\frac{\hbar ^{2}}{4\mu }\nabla ^{2}M\mathbf{n}.
\label{gforce}
\end{equation}%
The proof is as what follows.

In Heisenberg picture, the equation of motion for the momentum operator $%
\mathbf{\mathbf{p}}$ is, 
\begin{equation}
\frac{d}{dt}\mathbf{p}=\frac{1}{i\hbar }[\mathbf{p},H],  \label{HE}
\end{equation}%
where the momentum $\mathbf{p}$ and Hamiltonian $H$ are, respectively \cite%
{jk,dacosta,fc,liu11,liu13-1,ikegami,liu16,liu17,liu13-2,liu17-2}, 
\begin{equation}
{\mathbf{p}}=-i\hbar ({\nabla _{S}}+\frac{{M{\mathbf{n}}}}{2}),\text{ \ }H=-%
\frac{\hbar ^{2}}{2\mu }\nabla _{LB}^{2}+V_{G}=\frac{p^{2}}{2\mu }-\frac{%
\hbar ^{2}}{4\mu }\left( n_{i,j}\right) ^{2},  \label{GM}
\end{equation}%
where $V_{G}$ is the well-established geometric potential \cite%
{jk,dacosta,fc,ikegami},%
\begin{equation}
V_{G}=-\frac{\hbar ^{2}}{4\mu }\left( n_{i,j}\right) ^{2}+\frac{\hbar ^{2}}{%
8\mu }{M}^{2}=\frac{\hbar ^{2}}{4\mu }(\frac{1}{2}M^{2}-\left(
n_{i,j}\right) ^{2}),  \label{GP}
\end{equation}%
and $\nabla _{LB}^{2}={\nabla _{S}}\cdot {\nabla _{S}}$ is the
Laplace-Beltrami operator which is the dot product of the gradient operator $%
{\nabla _{S}\equiv }\nabla _{N}-\mathbf{n}\partial _{n}$ on the surface $%
\Sigma ^{N-1}$ with $\nabla _{N}$ being usual gradient operator in $R^{N}$.
The commutators between different components of the momentum ${\mathbf{p}}$ 
\cite{liu17-2} are, 
\begin{equation}
\lbrack p_{i},p_{j}]=\frac{i\hbar }{2}\left(
(n_{j}n_{i,l}-n_{i}n_{j,l})p_{l}+p_{l}(n_{j}n_{i,l}-n_{i}n_{j,l})\right) .
\end{equation}%
One component of the Heisenberg equation (\ref{HE}) becomes,%
\begin{equation}
\lbrack p_{j},H]=\frac{1}{2\mu }[p_{j},p_{k}p_{k}]-[p_{j},\frac{\hbar ^{2}}{%
4\mu }\left( n_{i,l}\right) ^{2}]=\frac{1}{2\mu }\left(
[p_{j},p_{k}]p_{k}+p_{k}[p_{j},p_{k}]\right) -[p_{j},\frac{\hbar ^{2}}{4\mu }%
\left( n_{i,l}\right) ^{2}].  \label{pih}
\end{equation}%
The last commutator in the right hand side of Eq. (\ref{pih}) can be
simplified into,%
\begin{equation}
\lbrack p_{j},\frac{\hbar ^{2}}{4\mu }\left( n_{i,l}\right) ^{2}]=2i\hbar 
\frac{\hbar ^{2}}{4\mu }(n_{i,l}n_{i,l,j}-n_{j}n_{k}n_{i,l}n_{i,l,k}).
\label{pjfg}
\end{equation}%
The commutator $[p_{j},p_{k}p_{k}]$ can be decomposed into two parts, $%
F_{j}+G_{j}$, which are \cite{liu17-2}, respectively,%
\begin{eqnarray}
F_{j} &\equiv &\frac{i\hbar }{2}\left\{
n_{j,l}n_{k}p_{l}p_{k}+p_{l}n_{j,l}n_{k}p_{k}+p_{k}n_{k}n_{j,l}p_{l}+p_{k}p_{l}n_{j,l}n_{k}\right\} 
\\
&\overset{c.l.}{\Rightarrow }&\frac{i\hbar }{2}\left\{ {\mathbf{p}}\mathbf{%
\cdot }\nabla n_{j}\mathbf{n\cdot }{\mathbf{p}}+{\mathbf{p}}\mathbf{\cdot }%
\nabla n_{j}\mathbf{n\cdot }{\mathbf{p}}+{\mathbf{p}}\mathbf{\cdot }\nabla
n_{j}\mathbf{n\cdot }{\mathbf{p}}+{\mathbf{p}}\mathbf{\cdot }\nabla n_{j}%
\mathbf{n\cdot }{\mathbf{p}}\right\} =2i\hbar {\mathbf{p}}\mathbf{\cdot }%
\nabla n_{j}\left( \mathbf{n\cdot }{\mathbf{p}}\right) ,  \label{Fj} \\
G_{j} &\equiv &-\frac{i\hbar }{2}\left\{
n_{j}n_{k,l}p_{l}p_{k}+p_{l}n_{j}n_{k,l}p_{k}+p_{k}n_{j}n_{k,l}p_{l}+p_{k}p_{l}n_{j}n_{k,l}\right\} 
\\
&\overset{c.l.}{\Rightarrow }&-\frac{i\hbar }{2}\left\{ n_{j}{\mathbf{p}}%
\mathbf{\cdot }\nabla \mathbf{n\cdot }{\mathbf{p}}+n_{j}{\mathbf{p}}\mathbf{%
\cdot }\nabla \mathbf{n\cdot }{\mathbf{p}}+n_{j}{\mathbf{p}}\mathbf{\cdot }%
\nabla \mathbf{n\cdot }{\mathbf{p}}+n_{j}{\mathbf{p}}\mathbf{\cdot }\nabla 
\mathbf{n\cdot }{\mathbf{p}}\right\} =-2i\hbar n_{j}{\mathbf{p}}\mathbf{%
\cdot }\nabla \mathbf{n\cdot }{\mathbf{p,}}  \label{Gj}
\end{eqnarray}%
where $c.l.$ denotes the classical limit. Clearly, $F_{j}$ (\ref{Fj}) goes
to zero in classical mechanics for we have an orthogonality $\mathbf{n\cdot }%
{\mathbf{p}}=0$, while $G_{j}$ (\ref{Gj}) corresponds to the centripetal
force $-2n_{j}{\mathbf{p}}\mathbf{\cdot }\nabla \mathbf{n\cdot }{\mathbf{p}}$%
. The quantities $F_{j}$ and $G_{j}$ can be simplified into, respectively, 
\begin{eqnarray}
F_{j} &\equiv &-i\hbar ^{3}n_{i,l}n_{i,l,j}, \\
G_{j} &\equiv &-2i\hbar (n_{j}\mathbf{\mathbf{p}}\cdot \mathbf{\nabla 
\mathbf{n}}\cdot \mathbf{\mathbf{p}}+\mathbf{\mathbf{p}}\cdot \mathbf{\nabla 
\mathbf{n}}\cdot \mathbf{\mathbf{p}}n_{j})-i\hbar
^{3}(n_{i,l}n_{i,l,j}-2n_{j}n_{k}n_{i,l}n_{i,l,j,k}-n_{j}n_{i,i,l,l}).
\end{eqnarray}%
Substitution of $F_{j}$, $G_{j}$ and $[p_{j},\frac{\hbar ^{2}}{4\mu }\left(
n_{i,l}\right) ^{2}]$ (\ref{pjfg}) into Eq. (\ref{pih}) directly leads to
the result (\ref{gemforce}). \textit{Q.E.D.}

Three immediate remarks are in order: 1. The calculation is straightforward,
but no one expects such a result before because it breaks down the Ehrenfest
theorem that implies no additional quantum force at all. 2. It is customary
to resort to the operator-ordering in the expression of the centripetal
force $-\mathbf{n}\left( {\mathbf{p}}\mathbf{\cdot }\nabla \mathbf{n\cdot }{%
\mathbf{p}}\right) $ to produce the expected form of the geometric
potential. However, desperate attempts have been made during last three
decades and all are unproductive \cite{liu17,packet3}. 3. In interface
physics, the Laplacian of mean curvature refers to the surface part $\nabla
_{LB}^{2}M$ \cite{packet2} rather than the full one $\nabla ^{2}M$ in
geometric force (\ref{gforce}), and it is interesting to note the difference
in between is in fact the normal component of the gradient of the geometric
potential, 
\begin{equation}
\nabla ^{2}M=\nabla _{LB}^{2}M+\partial _{n}(\frac{1}{2}M^{2}-\left(
n_{i,j}\right) ^{2}).
\end{equation}

Now, we estimate the magnitude of the curvature induced geometric force. It
is significantly different from zero in the area where the mean curvature
alters dramatically. For a two-dimensional curved surface with mean
curvature $M\sim 1/a$, $\chi _{g}\sim -\hbar ^{2}/(\mu a^{3})$. For $\mu
\sim 10^{-30}kg$, $a\sim 10^{-8}m=10nm$, $f_{g}\sim 10^{-2}pN$. For a
spheroid $\left( x^{2}+y^{2}\right) /a^{2}+z^{2}/b^{2}=1$, $\nabla ^{2}M$
reaches its maximum $-\left( b^{2}-a^{2}\right) b/a^{6}$ at the top $%
(x,y,z)=(0,0,b)$ for the prolate spheroid and it reaches its maximum $\left(
b^{2}-a^{2}\right) \left( b^{2}+3a^{2}\right) /\left( 2ab^{6}\right) $ at
the equator for the oblate spheroid, and for spherical surface, $\chi _{g}=0$%
. For a torus with $R$ being the distance from the center of the tube to the
center of the torus and $r$ ($\prec R$) being the radius of the circular
tube, $\nabla ^{2}M=R(r+R\sin \theta )/(2r^{2}(R+r\sin \theta )^{3})$
reaches its maximum $-R(R-r)/(2r^{2}(R-r)^{3})$ when $\sin \theta =-1$ which
lines out the circumference of the inside circle (with radius $R-r$) of the
torus.

In conclusion, in quantum mechanics for the particle on the hypersurface
there is mean curvature driven force. Though this force has no classical
correspondence, but well-established in interface physics, which causes the
curvature driven surface diffusion.

\begin{acknowledgments}
This work is financially supported by National Natural Science Foundation of
China under Grant No. 11675051.
\end{acknowledgments}

\end{document}